# Minimum wage and manufacturing labor share: Evidence from North Macedonia

**Marjan Petreski**[1]

**Jaakko Pehkonen**[2]

**Abstract**

The objective of the paper is to understand if the minimum wage plays a role for the labor share of manufacturing workers in North Macedonia. We decompose labor share movements on those along a share-capital curve, shifts of this locus, and deviations from it. We use the capital-output ratio, total factor productivity and prices of inputs to capture these factors, while the minimum wage is introduced as an element that moves the curve off. We estimate a panel of 20 manufacturing branches over the 2012-2019 period with FE, IV and system-GMM estimators. We find that the role of the minimum wage for the labor share is industry-specific. For industrial branches which are labor-intensive and low-pay, it increases workers' labor share, along a complementarity between capital and labor. For capital-intensive branches, it reduces labor share, likely through the job loss channel and along a substitutability between labor and capital. This applies to both branches where foreign investment and heavy industry are nested.

**Conflict of interest statement:** On behalf of all authors, the corresponding author states that there is no conflict of interest.

[1] Corresponding author. marjan.petreski@uacs.edu.mk, University American College Skopje, North Macedonia; Finance Think – Economic Research & Policy Institute, Skopje; Partnership for Economic Policies, Canada.

[2] University of Jyväskylä, Finland

## 1. Introduction

Labor share – the amount from the national income that is allocated to workers – has been long the topic off the focus in scientific debates. Primarily, the constancy of labor share was taken as a "stylized fact of growth" (Kaldor, 1961) and did not trigger much attention among scholars. The last two decades have seen a renewed interest, triggered by the empirical contestation of the stylized fact: the labor share started declining. Notable authors populated the scant literature on the evolution and causes of the labor share decline in the advanced economies; starting from the earlier prominent contributions like Acemoglu (2003) and Blanchard and Giavazzi (2003), to recent contributions like Autor et al. (2017), Atkeson (2020) and Kehrig and Vincent (2021).

Empirical evidence is yet diverse, without a unison conclusion about the causes of the labor share decline. For example, Atkinson (2009) argues that a smaller labor share attenuates the translation of macroeconomic gains into gains in personal incomes of households, while Piketty (2013) documents an adverse association between rising capital share and rising inequality. The decline has likewise given rise to political debates. Based on the key notion of how the generated value added in a society is divided between workers and the owners of capital, trade unions across Europe used the declining labor share fact to argue against policies supporting wage moderation, while governments frequently used it as a case for increasing taxation of profits and capital gains.

Developing and transition economies have not been an exception from such trends, despite with some delays and larger heterogeneity (Guerriero, 2012). Particularly transition economies exited the central planning system at the beginning of the 1990s; worker has been key in that economic system, which secured that labor shares are high and constant, and income inequality is moderate to low. However, the processes of transition, including but not limiting to privatization of the state-owned capital, the influx of multinational companies, and the flexibilization of labor markets occurred concurrently with the decline in the labor share, with even higher intensity than in the advanced economies (OECD, 2015).

Transition countries are diverse group of economies. For example, those in Central Europe followed a pattern of fast transformation over the 1990s, which secured that they caught up the EU average already in the 2000s. The countries of Southeast Europe followed a thornier path, including with political and military setbacks over the 1990s, which prevented them achieving more satisfactory economic outcomes before the 2010s. North Macedonia belongs to this group. Together with its regional peers, North Macedonia started a so-called 'race to the bottom' over the 2010s. The attempt was to offer the idle labor at the global stage and attract multinational companies in the free industrial zones against a set of subsidies, which in turn not only increased labor demand, but also supported re-industrialization of the economy, i.e., the rebounding of investment in fixed capital and technology.

In parallel, governments in the late transitioners, including North Macedonia, introduced and increased minimum wages. In North Macedonia, the first-ever minimum wage was introduced in 2012, positioned at about 39% of the average wage for all industries, except for textiles, leather an apparel, at 30%. Since then, a combination of economic and political factors resulted in a minimum wage currently at the 59% of the average wage and at over 80% of the median wage, with existing further upward pressure primarily driven by government promises but presently also supported by the inflationary pressures. At the same time, the labor share has been in the decline, from about 30% of output in the 1990s to 20% nowadays, but the moderation in the last decade – under the extant minimum wage – is evident.

From the viewpoint of the labor share, the role of the minimum wage is unwarranted: it either exerted no influence, or its potentially positive effect has been compensated by the capital accumulation and the capital-augmented technological progress. The objective of our study is to understand the role of the minimum wage for the labor share in North Macedonia. We pursue the objective by accounting for the capital endowments and capital-augmenting processes, hence being able to understand if the role of the minimum wage for the labor share changes depending on the relationship between labor and capital.

The paper brings a couple of novelties in the sparse of the referent literature. First, it is among the few investigating the role of the minimum wage for the labor share. Namely, the investigation of the supply—side factors and, recently, of globalization for labor share has received more attention. Even when considering the role of institutions in a general sense, studies have focused on employment protection mechanisms, unemployment insurance system and tax policies, rather than on the minimum wage per se. Second, the study is the first on transition economies to examine the minimum wage in the context of labor share. The scarce transition-economies literature on the minimum wages usually treats the employment and living-standard effect, but not the labor share explicitly. This may be important for global readership because transition countries may have exhibited a distinctive combination of capital-augmenting factors and insistence on minimum wages, which may serve specific policy lessons applicable elsewhere. Lastly, the study dwells attention by disaggregating industrial branches, which has never been done – to our knowledge – in the literature on developing economies.

The paper is structured as follows. Section 2 offers brief stylized facts on the labor share and the minimum wage. Section 3 lays the theoretical foundations and Section 4 presents the empirical derivation. Section 5 puts the empirical framework in the context of the existing literature, and Section 6 discusses the data and the estimation method. Section 7 presents the results. Section 8 concludes.

## 2. Brief stylized facts

The minimum wage in North Macedonia was first introduced in 2012, positioned at about 39% of the average wage for all industries, except for textiles, leather an apparel, at 30% (Figure 1). It covers both the private and public sector workers, with the exception of the self-employed workers. This minimum wage floor applies to the basic wage of a worker (with a full-time contract). Hence, in practice, workers' basic wage should not be lower than the minimum wage, without taking into consideration any wage supplements (either for good performance or in other circumstances, such as overtime work, night work, etc.).

Despite initially set to adjust with prices bi-annually, the rule was soon abandoned, as the minimum wage attained strong political power. Not long after its introduction, political parties commenced a 'battle' during every election to promise a higher minimum wage. This resulted in a de-facto full abandonment of the rule for adjustment and moved the discussion about the minimum wage level within the social dialogue of the country, considering also the articulation of the government to fulfill the pre-election promise (ILO, 2021). Hence, the minimum wage continued to be negotiated within the tri-partite social dialogue, comprised of the employers, trade unions and the government. By 2016, the net minimum wage increased by 25%, with reducing gap in the textiles, leather and apparel branches.

The elections of 2016 brought a leftist government in office, which made the minimum wage a central economic issue. Soon after the new government took office in 2017, the minimum wage rose by almost 30% and equalized in the textile, leather and apparel branches, for the first time somehow circumventing the social-dialogue table (Petreski et al. 2019). Then, the minimum wage continued to be a part of a collective bargaining process, whereby the government continues voicing its promises.

After a short stall during the pandemic period, the promise of the July 2020 elections to raise it at 18,000 MKD, was ultimately realized in March 2022, reflecting a near 20% increase for the second time, hence pushing it at about 58% of the average wage at the time. This occurred in parallel with the change of the adjustment mechanism of the minimum wage with both prices and the average wage growth, as well a prohibition that it falls lower than 57% of the average wage. It likewise happened in times when the inflationary shock from the global developments related to Ukraine started transmitting into the domestic economy, which meant that the higher minimum wage cushioned some of the living standard deterioration, but it might have contributed to the inflationary pressure itself; the recent estimate of Ramadani and Unevska-Andonova (2022) suggested that about 11% of the inflationary pressure in 2022 has been domestically borne. With the change in the adjustment formula, the minimum wage further increased to 20.174 in March 2023, representing 59% of the end-2022 average wage.

**Figure 1 – The Minimum Wage in North Macedonia**

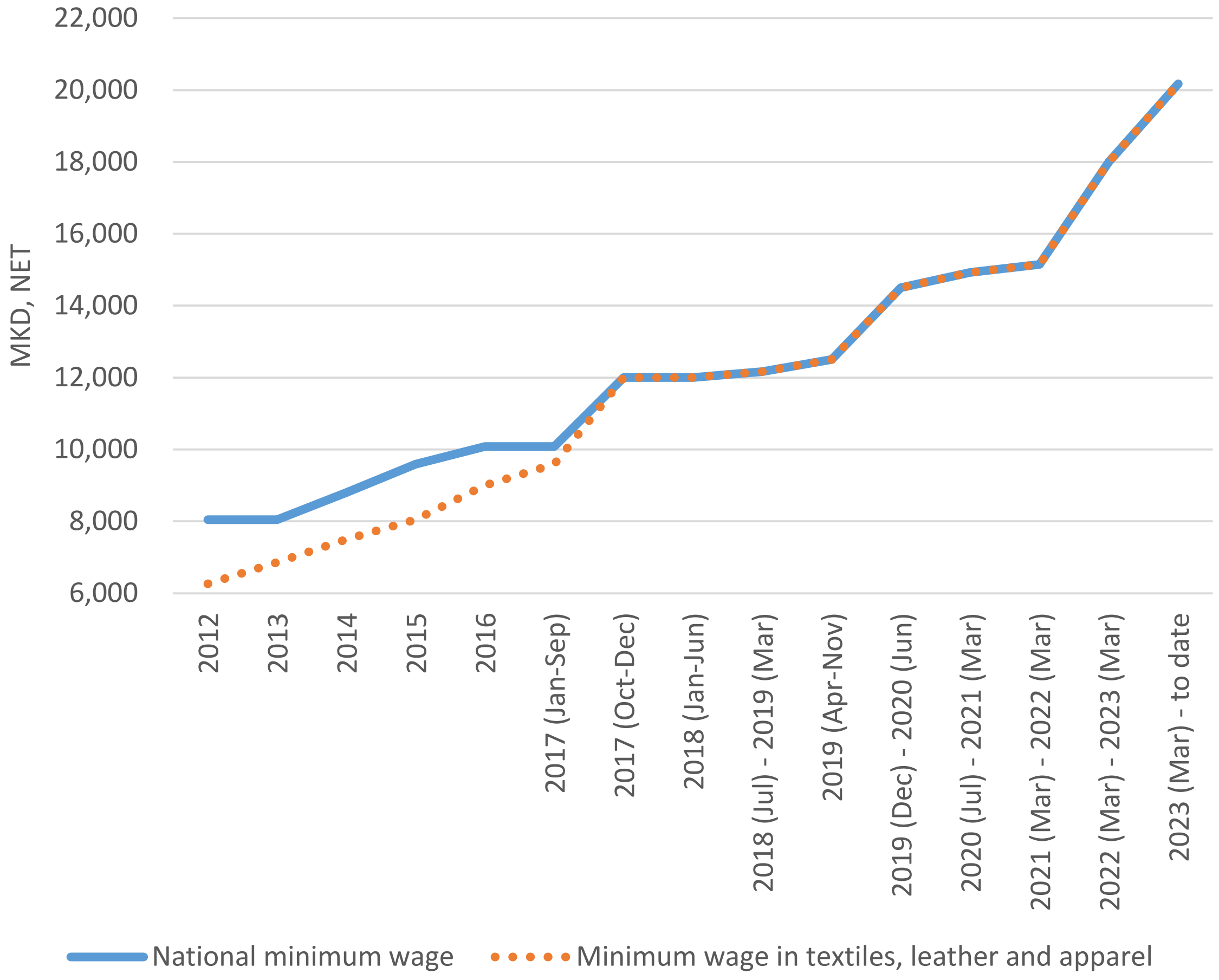

*Source: Minimum Wage Law.*

The labor share (in manufacturing) – here presented at Figure 2 – has been on a decline between 2000 and 2019, from about 30% to less than 20% of output. However, a moderation in the 2010s is evident, and particularly after the introduction of the minimum wage in 2012, at around 18% of the output. This legitimately opens the question if the minimum wage had a role to play for the labor share in North Macedonia – a question we answer in the following sections.

**Figure 2 – The Manufacturing Labor Share in North Macedonia**

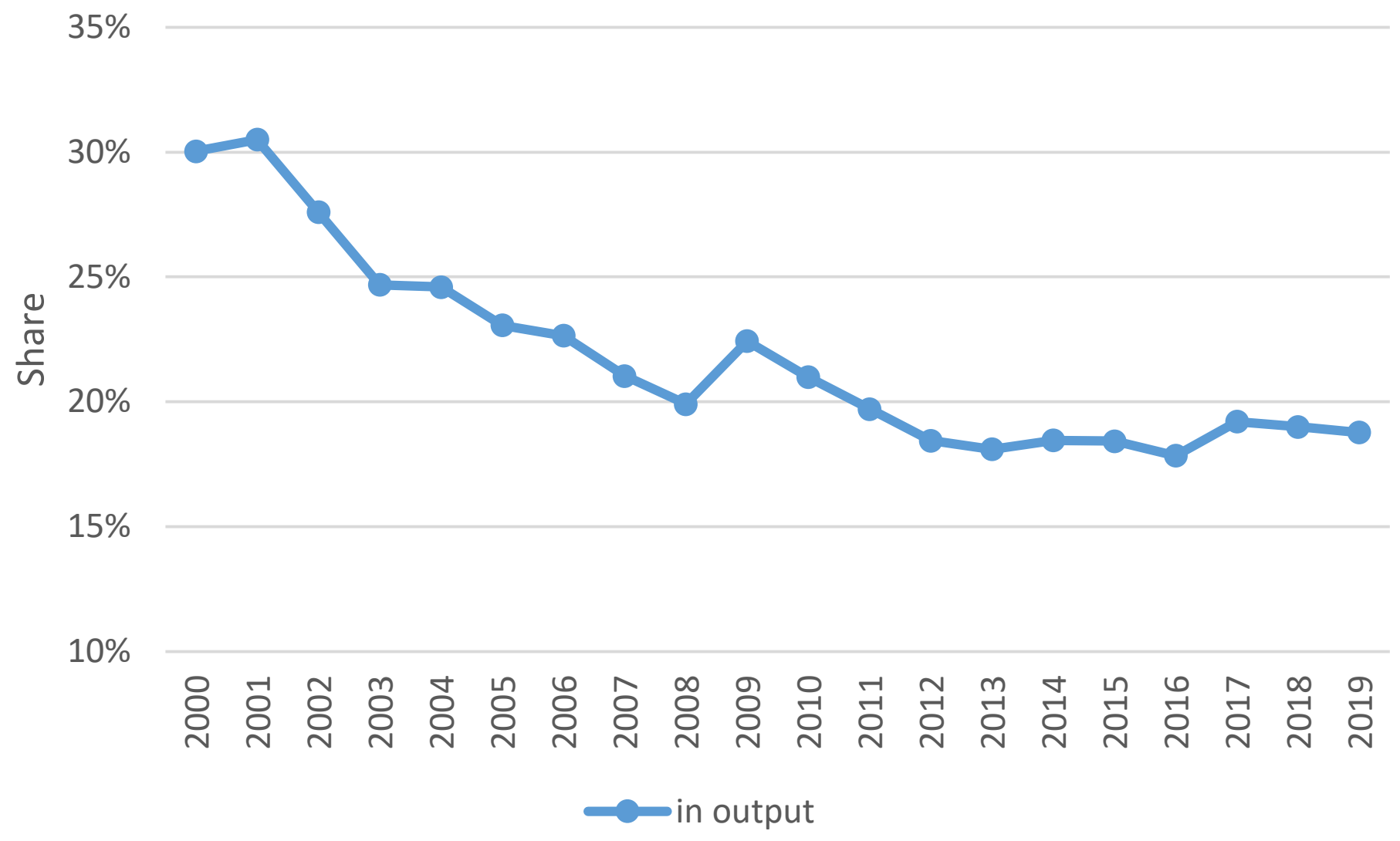

*Source: UNIDO Statistical Database.*

## 3. Theoretical basis

We devise the theoretical framework similar to the one in Bentolila and Saint-Paul (2003), by specifying a production function whereby the output $Y_i$ is produced with two factors of production, capital $K_i$, and labor $L_i$, and a standard labor-augmenting technological progress $B_i$:

$$Y_i = F(K_i, B_i L_i) \quad (1)$$

Then, under the assumption that the labor is paid the marginal product, there exists a unique function $g$ such that

$$s_{Li} = g(k_i) \quad (2)$$

whereby $s_{Li}$ is the labor share in industry $i$ and $k_i$ is the capital-output ratio $k_i = K_i/Y_i$, and equals

$$s_{Li} = \omega_i L_i / p_i Y_i, \quad (3)$$

whereby $\omega_i$ is the wage and $p_i$ is the product price.

Equations (2) and (3) reflect the stable relationship between the labor share and the capital-output ratio, denoted as the share-capital curve (SK curve). It implies that whenever factor prices (e.g. wages)  or quantities (e.g. number of workers) change, a move along the SK curve occurs, suggesting that a residual in (1) is not explained by these factors.

However, (2) and (3) would not hold if the technological progress is capital-augmenting, in which case we have:

$$Y_i = F(A_i K_i, B_i L_i) \quad (4)$$

which implies that the relationship between the labor share and the capital-output ratio is no longer stable, i.e., changes in $A_i$ may shift the SK relationship. If we assume a constant elasticity of substitution (CES), then we have:

$$Y_i = (\alpha(A_i K_i)^{\varepsilon} + (1-\alpha)(B_i L_i)^{\varepsilon})^{1/\varepsilon} \quad (5)$$

whereby $\varepsilon$ is also a technological parameter. Then, the labor share and the capital-output ratio are equal to:

$$s_{Li} = \frac{(1-\alpha)(B_i L_i)^{\varepsilon}}{\alpha(A_i K_i)^{\varepsilon}+(1-\alpha)(B_i L_i)^{\varepsilon}} \quad (6)$$

$$k_i = (\frac{K_i^{\varepsilon}}{\alpha(A_i K_i)^{\varepsilon}+(1-\alpha)(B_i L_i)^{\varepsilon}})^{1/\varepsilon} \quad (7)$$

It follows from (6) and (7) that:

$$s_{Li} = 1 - \alpha(Ak_i)^{\varepsilon} \quad (8)$$

which implies that the relationship between labor and capital is monotonic in $k_i$, i.e. either increasing or decreasing depending on the sign of $\varepsilon$. If $\varepsilon$ goes to zero, the production function converges to a standard Cobb-Douglas one. While for $\varepsilon < 0$, higher capital intensity reduces the labor share, hence labor and capital are substitutes, while for $\varepsilon > 0$, they are complements, suggesting that more capital intensity is associated with increasing labor share.

To introduce the minimum wage as a factor for labor share, we need to note that our SK curve, presented with equation (2), displays the relationship between the capital-output ratio and the employment elasticity of output (the labor intensity of growth). Hence, if the marginal product of labor – given with $s_{Li}$ – equals the real wage, the economy is on the curve. Two types of deviations are possible: a shift of the curve, like the one caused by the capital-augmenting technological progress described; or movements off the curve, which imply a difference between the marginal product of labor and real wage. We are interested in the latter, and particularly in how setting the minimum wage may fit into the framework.

To understand this, we revert to the bargaining model, as described in Bentolila and Saint-Paul (2003) where firms and unions bargain over the wages and then firms set employment. Because firms set employment given the pre-set wage level, the marginal product equation (2) remains valid. This implies that any changes to the minimum wage will cause a move along the SK curve, in a direction which depends on the elasticity of substitution between labor and capital. However, if firms and unions bargain over both wages and employment, then the marginal product of labor equals the real opportunity cost. In this case an increase in the bargaining power of workers increases labor share in the short run, but is not reflected in employment. In the long run, capital stock adjusts so that higher bargaining power of workers also triggers increases in employment. Then, out SK schedule assumes the following shape:

$$s_{Li} = \theta + (1-\theta)g(k_i) \quad (9)$$

whereby $\theta$ is the workers' bargaining power. It implies that the SK curve moves to right, the labor share increases, i.e., workers are paid more than their marginal product. The relationship infers that the higher the workers' bargaining power, the lower the sensitivity of the labor share to the capital-output ratio.

## 4. Empirical derivation

For empirical purposes, we assume a general multiplicative form of the labor share equation (2):

$$s_{L,it} = g(k_{it}, S_{it})h(X_{it}) \quad (10)$$

whereby the sub-indices denote industries $i$ and time $t$. $g(k_{it})$ captures the original SK schedule affected by S, which contains a measure of the total factor productivity and manufacturing real prices of inputs to capture the capital-augmenting technological progress. In North Macedonia, we assume this is the process led by the establishment of the free economic zones and the intensified influx of foreign direct investments (FDIs) following the campaign "Invest in Macedonia" over the 2010s. $h(X_{it})$ is meant to capture the discrepancy between the marginal product of labor and the wage – in our case the introduction and increase of the minimum wage. It could move the economy off the SK curve. Theoretically, we need a representative of $\theta$ – the bargaining power of workers in $h(X_{it})$ and due to the research question in this paper, we use the logarithm of the minimum wage. Certainly, bargaining power may have other more direct measures, like the unionization rate or the number of labor disputes (Bentolila and Saint-Paul, 2003), but on top of them being not of our interest in this study, the data frequently lack in North Macedonia; some information is available for the trade unionization rate, but it is neither industry-specific nor it exhibits yearly variations to be suitable for robust analysis.

As in Bentolila and Saint-Paul (2003), we assume that functions $g(k_{it}, S_{it})$ and $h(X_{it})$ are multiplicative:

$$g(k_{it}, S_{it}) = A_{it}^{\beta_0}(k_{it})^{\beta_{1i}}(\frac{q_{jt}}{p_{jt}})^{\beta_{2i}} \quad (11)$$

$$h(X_{it}) = \exp(\beta_3 x_{it}) \quad (12)$$

where $A_{it} = TFP_{it}$, the total factor productivity, $\frac{q_{jt}}{p_{jt}}$ stands for the price of imported raw materials, $X_{it} = (\ln mw_{it}, v_{it})$, whereby $\ln mw_{it}$ is the log of the real minimum wage in place, with subscript $i$ to reflect it was industry-wise at least in part of the observed period, while $v_{it}$ is the residual that contains other factors that could move off the economy from the SK schedule, such as markups, hiring and firing costs, of which some are hard to measure. If we substitute (11) and (12) in (10) and take logs, we obtain the estimable equation:

$$\ln s_{L,it} = \beta_0 \ln TFP_{it} + \beta_1 \ln k_{it} + \beta_2 \ln (q_{it}/p_{it}) + \beta_3 \ln mw_{it} + v_{it} \quad (13)$$

## 5. Discussion from the perspective of existing literature

Our estimable model (13) is nested into the empirical labor-market institutions literature where the dependent variable in a reduced-form equation – the labor share in our case – is set to depend on three factors: supply-side shocks, institutions, and demand-side controls. Prominent articles in this regard include Nickell (1997), Daveri and Tabellini (2000), Elmeskov et al. (1998), Blanchard and Wolfers (2000), Nickell et al. (2005) and Bentolila and Jimeno (2006). Certainly, in this strand of literature, labor-market institutions are treated widely, from those capturing employment and unemployment protection to collective bargaining, hence not exclusively focusing on the minimum wage. For example, Young and Zuleta (2018) analyzed how union density interferes with the labor shares using panel data on 35 US industries between 1983 and 2005, and document that stronger unions are associated with increasing labor shares. Guschanski and Onaran (2022) found that the wage share in 14 OECD countries declined due to a fall in labor's bargaining power driven by offshoring to developing countries and changes in labor market institutions such as union density, social government expenditure and minimum wages. Bentolila and Saint-Paul (2003) hardly find industry-differentiating effects of the bargaining power for the labor share in 12 OECD countries. We refrain from reviewing empirical studies for single advanced economies, as they may be a subject of a separate meta-analysis due to their volume.

We are, yet, interested in one the institutions of the labor market: the minimum wage, which we position as an outcome of a collective bargaining process, which has been rarely done in the literature. Empirical studies using the level of the minimum wage include, e.g., ILO (2012); IMF (2007). OECD (2012) provides empirical evidence that the roles of the factors that affect the bargaining power of workers is the largest. However, the empirical literature that explicitly relates the minimum wage with the labor share is so scant that we cannot offer any comparative estimate that may motivate our own discussion. Moreover, the literature that combines developing-economy and industry-level estimates is almost inexistent. Harasztosi and Lindner (2019), for Hungary, touch upon some of the issues we consider in this study and find that firms responded to the minimum wage increases by substituting labor for capital as well by passing the burden onto consumers through prices. In cases when such passing proved more difficult, a grater disemployment effects were registered. However, they did not analyze the role for labor share per se.

Supply side is captured through shocks in oil prices, terms of trade, and productivity (or total factor productivity) (Judzik and Sala, 2013). Technological changes are often presented as the key culprit, with an abundancy of literature seeing capital accumulation and the capital-augmenting technological progress as key determinants of the labor share, which is fully aligned with our theoretical model. Papers include: Arpaia et al. (2009); Driver and Muñoz-Bugarin (2010); Raurich et al. (2012); Karabarbounis and Neiman (2014); Broman (2021). In our case, we use two variables to capture the supply-side: the total factor productivity and capital-output ratio.

The demand-side factors for labor share have been relegated to the minimum in the literature. However, they have received more attention recently, particularly in developing economies. Namely, the increased exposure of the economies to the process of globalization, put workers in the nexus so as to understand if international trade and, particularly, the accelerated influx of multinational companies, resulted in shrinking welfare of workers amid the ‘race to the bottom’ supported by tax exemptions and generous subsidies (Petreski, 2021). In the scant literature discussing the role of globalization for labor share (Rodrik, 1998; Harrison, 2005; Lee and Jayadev, 2005; Doan and Wan, 2017; Tytell and Jaumotte, 2008; Autor et al. 2020), economic openness and FDI inflow are usually used. In our model, however, to be consistent with our theoretical derivation, we include the prices of manufacturing inputs in the domestic production process. This variable could be contested from the viewpoint of solely capturing the demand side, because it could also reflect the capital-augmenting technological progress, to the extent it reflects the import conditions of raw materials. It, however, cannot be denied that, for countries like North Macedonia, it would also capture the prevailing conditions in international trade, given the large exposure of the manufacturing industry to imports.

## 6. Data and estimator

We use data from the UNIDO Industrial Database, providing us with information for manufacturing industrial branches at the two-digit ISIC classification, for each of the years between 2012 and 2019. The reasons for relying on this period are manifold: first, the minimum wage, which is of central interest here, has been introduced in 2012; second, the UNIDO database has a gap for 2011; third, it is the period after the global financial and European sovereign crises. So, from multiple angles, the period before 2012 exhibits a structural break whose modelling is beyond the scope of this study.

We operate with 20 branches of manufacturing industry (out of the total of 22), which provides a dataset of 160 observations and a balanced panel. Data and variables’ descriptions are provided in Annex 1.

The labor share is constructed by dividing the amount paid as wages and wage supplements to workers with the total industrial output. The capital share is the ratio of the gross fixed capital formation per industrial branch and the total output. The total factor productivity is a derived variable in a standard growth accounting exercise in which the output, number of workers and fixed capital formation per industry feature as inputs.

Prices of imported raw materials are approximated by the prices of inputs in domestic production, which is sourced from the price statistics of the State Statistical Office of North Macedonia. The variable does not exactly capture import prices, because such are not published at the industrial level, but should approximate both capital-augmenting technological progress and some facets of globalization to the extent industrial branches are exposed to import.

The variable on the minimum wage is constructed from the Law on the Minimum Wage adopted in 2012 and its subsequent amendments implying minimum wage increases.

Equation (13) is estimated using panel data techniques. We commence the estimation with a fixed-effects estimator, which provides a baseline description of the relationships we would like to understand. In particular, with a simple approach, we are able to differentiate some of the effects by industry branch. One may argue, however, that some endogeneity is present in our relationship, particularly on the link between labor share and the variables capturing capital endowment and capital-augmented technological progress. This is implicitly present in our theoretical model where both factors of production are determined in a single framework in which they could be substitutes or complements. Although import prices are given for a small economy as North Macedonia is, prices on inputs in the domestic production process may suffer weak exogeneity. Similarly, the minimum wage is a result of a tripartite dialogue, whereby the interference with workers' incomes is a non-negligible topic of consideration, but the imposition of the minimum wage level by the government as a pre-election topic introduces exogenous elements in it.

Consequently, we treat labor share, the capital-output ration, the total factor productivity, the input prices, and the minimum wage as potentially endogenous. We use the information contained in the lagged values of the suspected endogenous variables as instruments. First, we rely on a standard instrumentation where the lags in levels are used to instrument current values of the endogenous variables. Second, we rely on an Arellano-Bover (1995) system GMM estimator, which introduces a dynamic regressor in the estimation – the lagged dependent variable, and which is shown to yield potentially large efficiency gains vis-à-vis the pure difference estimator. In a system GMM estimator, errors in levels are instrumented through the first differences, and vice versa. We rewrite equation (13) as follows:

$$y_{it} = \beta' x_{it} + v_{it} \qquad (14)$$

whereby: $y_{it} = \ln s_{L,it}$, $x_{it} = (\ln TFP_{it}, \ \ln k_{it}, \ln(q_{it}/p_{it}), \ln mw_{ij})$ and $\beta$ is the parameter vector. The $v_{it}$ are assumed to be independently distributed across units with zero mean, but arbitrary forms of heteroskedasticity across units and time are allowed. The identification is as follows. If there is a variable, $Z_{it}^{L}$ which satisfies the condition $E(Z_{it}^{L}\epsilon_{it}) = 0$, and we can assume that $E(Z_{it}^{L}\delta_i)$ does not depend on *t,* then we have $E(\Delta Z_{it}^{L} v_{it}) = 0$, i.e. $\Delta Z_{it}^{L}$ is a valid instrument for equation (14). Similarly, for the equation estimated in first differences, $E(Z_{it}^{D}\Delta v_{it}) = 0$ implies that $Z_{it}^{D}$ is a valid instrument. With this, the estimated equation provides information on the determinants of the level of labor share, i.e., its variations across industries with the changes in the other regressors.

For the estimations based on instrumental variables, we report the Hansen test statistics which tests the overidentifying restrictions for the validity of the instrument set.

## 7. Results

### 7.1. Baseline results

Table 1 presents the baseline estimates. Three general findings should be noted. First, the estimates are similar across estimators. Second, the results show that the null hypothesis of instruments validity cannot be rejected. Third, for the standard IV estimation, the results fail to reject the null that suspected endogenous regressors could be treated as exogenous.

The capital-output ratio is positively related with the labor share, suggesting that capital and labor are complements. An increase of the capital share by one percent is associated with an increase of the labor share by 0.2 percent. Such a result can be expected given the inflow of multinational companies in the last decade: they contributed to a significant increase of the capital formation in the country as well created a strong demand for labor, both of which result in a positive relation. The estimate for total factor productivity is negative, statistically significant and different than the one for the capital-output ratio. The results suggest that TFP is not strictly capital-augmenting.

The key variable of interest – the minimum wage – is insignificant across specifications, documenting no role of the minimum wage for the labor share. The result suggests that the insistence on fast minimum wage increases not necessarily translate into higher income of workers.

**Table 1 – Baseline results**

| | FE | IV | | System GMM | |
|---|---|---|---|---|---|
| | | All endogenous | KO and TFP endogenous | All endogenous | KO and TFP endogenous |
| | (1) | (2) | (3) | (4) | (5) |
| **Log labor share lagged** | | | | 0.612*** | 0.535*** |
| | | | | (0.047) | (0.064) |
| **Log capital-output ratio** | 0.221*** | 0.163** | 0.199*** | 0.113** | 0.0873* |
| | (0.021) | (0.067) | (0.068) | (0.049) | (0.051) |
| **Log TFP** | -0.092*** | -0.120** | -0.0325 | -0.204*** | -0.203*** |
| | (0.020) | (0.056) | (0.067) | (0.043) | (0.040) |
| **Log prices of inputs** | -0.518 | -0.609 | -0.211 | -0.163 | -0.121 |
| | (0.322) | (0.855) | (0.181) | (0.350) | (0.423) |
| **Log minimum wage** | 0.0788 | -0.114 | -0.00061 | -0.0878 | -0.123 |
| | (0.096) | (0.085) | (0.125) | (0.090) | (0.115) |
| **Constant** | 1.312 | | | 3.253 | 3.131 |
| | (1.462) | | | (2.122) | (2.716) |
| | | | | | |
| **Observations** | 160 | 120 | 120 | 140 | 140 |
| **Number of cross-sections** | 20 | 20 | 20 | 20 | 20 |
| | | | | | |
| **R-squared** | 65.7% | 24.2% | 24.3% | - | - |
| | | | | | |
| **Hansen test of overid. restrictions (p-value)** | - | 0.4158 | 0.3408 | 0.5270 | 0.2520 |
| Ho: Instruments are valid instruments | | | | | |
| **Endogeneity test (p-value)**<br>Ho: The specified endogenous regressors can be treated as exogenous | - | 0.5653 | 0.6903 | - | - |

*Source: Authors' estimations.*
**, ** and *** signify a statistical significance at the 10%, 5% and 1% level, respectively. Standard errors given in parentheses. Time dummies included to control for economy-wide shocks over time. Optimal lag length based on the Akaike information criterion, where applicable.*

However, the result may reflect aggregation across manufacturing branches whose sensitivity to the minimum wage increases vary. Namely, the manufacturing industry of North Macedonia is quite diverse and spanning from low-productivity industries like textiles and apparel (whose weakness was recognized even through the lower-than-the-national minimum wage introduced in 2012) to well-established domestically-owned exporters (like in pharmaceuticals) to technologically advanced branches, like automotive industry, located in the free zones. Hence, we continue for exploration of such differential effects.

### 7.2. Differential effects by industry

Table 2 presents the disaggregated results using the FE-estimator, as IV-based results are not possible to be produced under many cross-terms which require massive number of instruments. We present the interaction terms for each of the four variables one by one, as their joint estimation results in a sizeable loss of degrees of freedom. Branch capital-output ratios are jointly very significant (p-value = 0.000), which confirms the departure from the Cobb-Douglas production function. Negative coefficients (among the significant ones) dominate, with the exception of the textiles and apparel branches. This suggests that, except in these two, labor and capital are substitutes. In the textiles and apparel branches, the positive sign on the capital-output ratio suggests they are complements, which may well correlate with the low technological level in these branches, as well as the (unobserved) branch shares of skilled labor.

Branch-disaggregated TFP effects are also jointly significant (p-value = 0.0001), with varied signs, as well as signs which are predominantly opposite of the signs on the capital-output ratio. This suggests that TFP is not strictly capital-augmenting, corroborating our general finding. This is the case for all branches, except textiles and paper production, suggesting that a more complex effect of productivity on the production function may be at work there.

The disaggregation of the input prices – one of the shifters of the SK schedule – does not lead to further insights: some interactions are individually significant, but their joint significance cannot be warranted as in the aggregated case (p-value = 0.145).

The opposite holds true for the minimum wage, which brings interesting and important insights and conclusions. The insignificance of the aggregated result (Table 2) is converted into significant disaggregated effects in 16 out of 20 industrial branches (and high joint significance, p-value=0.000), and with a significant variety.

In food and beverages, textiles, apparel, coke and other transport equipment, an increase of the minimum wage is associated with larger share that goes to workers. This is an important finding for at least two reasons. First, these five branches produce 25% of the total manufacturing output and employ half of the manufacturing workers (50.1%, 2019 figures) and, second, they are dominated by small and medium-sized enterprises, predominantly of domestic ownership. In this group, the textiles and apparel branches deserve special attention. They employ a significant share of workers at the minimum wage, about 50% (Petreski et al. 2019). They are also the only two branches whereby labor and capital were assessed as complementary factors of production. In conjunction – the complementarity and the increasing labor-share effect of the minimum wage – suggest that increasing minimum wage policy and any attempt (including government subsidies) to elevate the technological levels are unlikely to exacerbate workers' income; quite the contrary, the minimum wage and capital investment could help the branches to escape the low-productivity – low-wage trap. This may further reinforce the role of TFP in textiles, which is found to play strictly capital-augmenting role.

This positive story has its adversary, however. In 11 manufacturing branches, the increase of the minimum wage is associated with a reducing labor share, documenting a potential negative impact in these industries. Contrary to those above, these industries employ only a bit more than quarter of the manufacturing workers (27.7%, 2019), but they generate 61.4% of the manufacturing output. Two subgroups deserve special attention. The first is the electrical machinery and apparatus, and chemicals branches, large part of which are constituted of factories – large employers - operating in the free economic zones. The second subgroup is the basic metals and non-metallic products, which is representing the old Macedonian industry, part of which has been later sold to foreign investors (outside the free economic zones). Hence, in part, these branches depend on the global developments with metal prices. In both subgroups, labor and capital are found substitutes, which suggests that higher minimum wages may be borne with intensification of the capital base at the expense of the number of workers, which in turn reduces the labor share, along the findings of e.g. Harasztosi and Lindner (2019).

**Table 2 – Disaggregated results, by manufacturing branch**

| | **Disaggregation by:** | | | |
|---|---|---|---|---|
| | **Capital-output ratio** | **TFP** | **Input prices** | **Minimum wage** |
| | (1) | (2) | (3) | (4) |
| Log capital-output ratio | | 0.241*** | 0.182*** | 0.216*** |
| | | (0.050) | (0.044) | (0.034) |
| Log TFP | -0.067*** | | -0.0968*** | -0.0711*** |
| | (0.016) | | (0.024) | (0.023) |
| Log prices of inputs | -0.478 | -0.434 | | -0.183 |
| | (0.308) | (0.419) | | (0.215) |
| Log minimum wage | 0.0793 | 0.0894 | 0.001 | |
| | (0.078) | (0.108) | (0.102) | |
| | | | | |
| 15 Food and beverages | 0.421 | -0.0760*** | -2.028*** | 0.175*** |
| | (0.302) | (0.012) | (0.687) | (0.013) |
| 16 Tobacco | 0.473 | 0.104*** | -0.00824 | -0.355*** |
| | (0.326) | (0.025) | (0.916) | (0.036) |
| 17 Textiles | 0.754** | 0.104*** | 6.020*** | 0.550*** |
| | (0.353) | (0.013) | (1.509) | (0.047) |
| 18 Wearing apparel | 0.489* | -0.0403** | -2.139*** | 0.525*** |
| | (0.263) | (0.018) | (0.359) | (0.036) |
| 19 Leather | 0.386 | 0.0539 | 1.900** | -0.0117 |
| | (0.262) | (0.061) | (0.667) | (0.018) |
| 20 Wood products | -0.278 | -0.0287*** | 2.902* | -0.0963*** |
| | (0.399) | (0.010) | (1.656) | (0.021) |
| 21 Paper | -1.389*** | -0.0380* | 3.664*** | -0.494*** |
| | (0.183) | (0.022) | (0.312) | (0.021) |
| 22 Printing | -0.618* | 0.00428 | 2.538** | 0.04 |
| | (0.334) | (0.013) | (1.210) | (0.038) |
| 23 Coke and petroleum | -0.200 | 0.0565 | -0.735 | 0.575*** |
| | (0.292) | (0.111) | (1.190) | (0.186) |
| 24 Chemicals | -0.429* | 0.0568*** | 5.867*** | -0.205*** |
| | (0.255) | (0.015) | (1.412) | (0.016) |
| 25 Rubber and plastics | -0.0453 | -0.0228 | 2.437* | -0.133*** |
| | (0.225) | (0.016) | (1.249) | (0.037) |
| 26 Non-metallic products | -1.166*** | -0.0012 | 3.884*** | -0.611*** |
| | (0.160) | (0.012) | (0.464) | (0.021) |
| 27 Basic metals | -0.483* | 0.102*** | 0.931 | -0.468*** |
| | (0.255) | (0.027) | (0.818) | (0.050) |
| 28 Fabricated metal products | -0.198 | -0.0126 | 2.155** | -0.353*** |
| | (0.354) | (0.018) | (0.851) | (0.043) |
| 29 Machinery and equipment | 0.293 | -0.176*** | 3.908*** | -0.878*** |
| | (0.338) | (0.030) | (0.707) | (0.061) |
| 30 Office and computing machinery | -0.121 | -0.0888** | 1.139 | -0.312** |
| | (0.312) | (0.040) | (0.831) | (0.116) |
| 31 Electrical machinery and apparatus | -0.530* | -0.0126 | 1.510* | -0.554*** |
| | (0.262) | (0.012) | (0.825) | (0.046) |
| 32 Media and communication equipment | -0.21 | 0.0699 | 1.551 | -0.136 |
| | (0.309) | (0.062) | (8.667) | (0.199) |
| 35 Other transport equipment | 0.368 | -0.102*** | 5.008*** | 0.111*** |
| | (0.298) | (0.009) | (1.529) | (0.026) |
| 36 Furniture | -0.282 | 0.0456*** | 2.693** | 0.0251 |
| | (0.287) | (0.008) | (1.110) | (0.052) |
| | | | | |
| F-test (p-value)<br>Ho: All interactions are jointly significant | 0.0000 | 0.0001 | 0.1450 | 0.0000 |
| Constant | 0.881 | 0.701 | -1.452 | -0.07 |
| | (0.944) | (1.808) | (1.768) | (0.973) |
| | | | | |
| Observations | 160 | 160 | 160 | 160 |
| R-squared | 86.4% | 70.3% | 72.7% | 78.5% |
| Number of cross sections | 20 | 20 | 20 | 20 |

*Source: Authors' estimations.*

**, ** and *** signify a statistical significance at the 10%, 5% and 1% level, respectively. Standard errors given in parentheses. Time dummies included to control for economy-wide shocks over time. Optimal lag length based on the Akaike information criterion, where applicable.*

### 7.3. Labor-intensive versus capital-intensive aggregations

Given these disaggregated results, we are able to come back to the IV and system-GMM estimates, to soften any critique that reliance on an FE estimator neglects any problem of endogeneity. Namely, we classify the manufacturing branches on labor-intensive (17-22, 36) and capital-intensive (15, 16, 23-35), following the ranking in Kucera and Sarna (2006), and pursue disaggregation of the minimum-wage effect on labor share based on such dichotomy. Table 3 presents the results. All estimates are similar with those in Table 1, both in terms of magnitude and significance. The separation of the minimum wage effect by factor intensity lends support to our findings in Table 2: the minimum wage increase leads to an increase of the labor share in labor-intensive branches, and the opposite is true in the capital-intensive branches. The latter result is, however, not sustained, as the significance is barely appearing, but it could be a result of the higher heterogeneity of branches within the capital-intensive group itself. On the other hand, the effect of the minimum wage in labor-intensive industries is robust. Considering endogeneity in the analytical framework (more so for the capital endowment than of the minimum wage itself), halves the estimate, while introducing dynamics in the framework reduces the estimate to about 0.04 (the long-run coefficient, calculated as _b[lmw]/(1-_b[l.lls]), is 0.074 and significant at 1%). Hence, both issues may be actually attenuating the manner in which the minimum wage works for labor share in labor-intensive manufacturing branches, but the evidence for the minimum wage being endogenous remains feeble.

**Table 3 – Disaggregated result, by factor intensity**

| | **FE** | **IV** | | **System GMM** | |
|---|---|---|---|---|---|
| | | All endogenous | KO and TFP endogenous | All endogenous | KO and TFP endogenous |
| | (1) | (2) | (3) | (4) | (5) |
| **Log labor share lagged** | | | | 0.452*** | 0.449*** |
| | | | | (0.096) | (0.054) |
| **Log capital-output ratio** | 0.223*** | 0.161*** | 0.197*** | 0.055 | 0.0660** |
| | (0.019) | (0.061) | (0.065) | (0.035) | (0.032) |
| **Log TFP** | -0.0882*** | -0.105* | -0.0278 | -0.166*** | -0.173*** |
| | (0.021) | (0.056) | (0.068) | (0.056) | (0.044) |
| **Log prices of inputs** | -0.524 | -0.412 | -0.209 | -0.37 | -0.0695 |
| | (0.319) | (0.862) | (0.179) | (0.528) | (0.310) |
| **Log minimum wage in capital-intensive branches** | -0.0562 | -0.167* | -0.0609 | -0.0509 | -0.141* |
| | (0.085) | (0.098) | (0.138) | (0.104) | (0.082) |
| **Log minimum wage in labor-intensive branches** | 0.391** | 0.185* | 0.196* | 0.0400** | 0.0409** |
| | (0.176) | (0.112) | (0.112) | (0.016) | (0.016) |
| **Constant** | 1.289 | | | 2.92 | 2.431 |
| | (1.213) | | | (2.848) | (2.086) |
| | | | | | |
| **Observations** | 160 | 120 | 120 | 140 | 140 |
| **Number of cross-sections** | 20 | 20 | 20 | 20 | 20 |
| | | | | | |
| **R-squared** | 68.9% | 29.8% | 25.9% | | |
| | | | | | |
| **Hansen test of overid. restrictions (p-value)** Ho: Instruments are valid instruments | | 0.3973 | 0.3915 | 0.1050 | 0.2130 |
| **Endogeneity test (p-value)** Ho: The specified endogenous regressors can be treated as exogenous | | 0.4406 | 0.6038 | | |

*Source: Authors' estimations.*

**, ** and *** signify a statistical significance at the 10%, 5% and 1% level, respectively. Standard errors given in parentheses.*

*The dummy which classifies the branches on labor- and capital-intensive is wiped out by the industry fixed effects. Time dummies included to control for economy-wide shocks over time. Optimal lag length based on the Akaike information criterion, where applicable.*

### 7.4. Robustness checks

An apparent critique related to the central variable of interest is the fact that we use its logarithm. Instead, the literature is abundant of utilization of the level of the minimum wage with regard to the average or the median wage (the so-called Kaitz ratio); examples include: Harasztosi and Lindner (2019); Aaronson et al. (2018). Additional advantage, in our case, of using a relative indicator of the minimum wage is the notion that the average wages per industrial branch will introduce further variation in out minimum wage variable. The results are presented in the Annex 2, mainly because they produce almost the same picture and conclusions from before. In the annex, the few deviations from the estimations based on the logged minimum wage are marked with italics. We pay some attention to them.

The most important change is that significant coefficients are obtained for the leather industry, while this was not the case before (Table A2.2). The positive capital-output ratio suggests that, similar to the textiles and apparel, the leather industry has both factors of production as complements, while the positive coefficient on the minimum wage adds that it plays a positive role for the labor share. It is not strange that the leather branch joins the group of textiles and apparel, the three braches which are known as low-productivity – low-pay and were initially introduced with a lower minim wage than the national one. Similarly, the minimum wage plays a positive role for the labor share in the printing branch, while was insignificant before, while the previously effect in rubber and other transport equipment branches is now found to be inexistent.

## 8. Conclusions and policy implications

In this paper, we aimed to understand if the minimum wage plays a role for the workers' shares in North Macedonia. As in the academic literature, the policy scene in the country is overwhelmed with the discussion of whether large promised minimum wage increases may actually exacerbate jobs rather than support the income of workers.

Our analytical framework consists of decomposition of labor share movements on those along a share-capital curve, shifts of this locus, and deviations from it. Movements along the curve comprehend changes in factor prices such as wage and real interest rate increases, as well the contribution of the labor-augmenting technological progress. The curve is shifted by factors like capital-augmenting technological progress or changes in import prices. Lastly, other sources of variation in the labor share are represented by movements off the curve, capturing the deviations from the marginal cost pricing, and here we nest the minimum wage as a representation of changes in workers' bargaining power.

We test the empirical power of this model on the case of North Macedonia's manufacturing industry. We use data of a panel of 20 manufacturing branches over the 2012-2019 period. We use FE estimator, IV, and system-GMM estimator, whereby former values of the included variables are used as instruments and their validity is tested through standard statistical tests.

We find evidence that the answer to our general question is not straightforward. The role of the minimum wage is strongly industry-specific. For industrial branches which are labor-intensive and low-pay – most notably textiles and apparel – the minimum wage plays a positive role for the workers' shares. The finding is concomitant to the complementary role of the capital endowment and labor, while in textiles, the production process is strongly capital-augmenting.

This finding suggests that workers in these branches would benefit from minimum wage increases only if these are accompanied by investments in higher-technology machines, equipment and processes. As a policy prescription, this is particularly relevant in the current context whereby the upward pressure onto the minimum wage currently generated by the inflationary pressures appended to the upward pressure generated from the government promises, which started as early as in 2016. In a context of dampening real output due to the consequences of the conflict in Ukraine, as well the insufficient competitiveness of the textile-apparel branches, may actually compromise the positive role of the minimum wage for the workers' shares, unless the sector is supported to elevate its production capacities, technology and innovation potential. Hence, particularly if the government decides to significantly reduce and wage subsidy (which was in place by the end of 2022), then the natural follow up – and a more sustained approach – is to offer a palette of supporting measures for purchase of equipment and supportive approaches for stronger integration of these branches into the global value chains, particularly for lifting up their position in these chains.

On the other hand, however, in a multitude of other branches, of which major part capital-intensive, the minimum wage increases negatively affect the labor share, likely working through the job loss channel. This is not strange, because the finding is accompanied by identified substitutability between labor and capital in these branches. This applies to both branches where FDIs are nested – electronics and chemicals, and where old heavy industry lays – metals and non-metallic products. Hence, if these branches are able to respond to the minimum wage increases with strengthening capital endowments and the technological processes instead of employing more workers, they will likely respond in such a fashion.

The policy advice related with this second strand of findings would be that, if the government further exerts minimum wage increases – unrelated to inflationary developments – which apparently jeopardize job generation in the capital-intensive industries, and with a potentially detrimental role for the attraction of new FDIs, then it may shift the focus in these branches from simple jobs generation to generation of smaller number but high-paid jobs. This is supportive to the current context whereby the Macedonian labor market faces intensifying labor shortages, which creates wage pressures themselves as well as supports the substitution from labor to capital wherever possible. This means that our finding supports a policy whereby high-paid jobs may be the primary objective to be attained by the means of the market forces, that way also making the minimum wage increases for these branches less and less relevant over time.

**Conflict of interest:** The authors declare no conflict of interest.

## Annex 1 – Data and variable descriptions

### Table A1.1 – Data descriptive statistics

| Variable | Obs | Mean | Std.Dev. | Min | Max |
|---|---|---|---|---|---|
| **Log labor share** | 160 | -1.86771 | 0.691347 | -4.23048 | -0.78112 |
| **Log capital-output ratio** | 160 | -2.5852 | 1.455203 | -5.59452 | 2.401196 |
| **Log TFP** | 160 | 10.16629 | 0.907465 | 7.766417 | 13.74218 |
| **Log prices of inputs** | 160 | 4.631376 | 0.05965 | 4.533674 | 4.925803 |
| **Log minimum wage** | 160 | 9.177535 | 0.177511 | 8.742414 | 9.434044 |

### Table A1.2 – Variable description

| Variable | Description | Source |
|---|---|---|
| **Labor share** | Log of the share of workers' compensation (wages and supplements) in branch output. | UNIDO Industrial Statistics |
| **Capital-output ratio** | Log of the share of gross fixed capital formation in branch output. | UNIDO Industrial Statistics |
| **TFP** | Log of the residual in a regression of the output on the number of employees and the gross capital formation. | UNIDO Industrial Statistics |
| **Prices of inputs** | Log of prices of inputs in the domestic market (index, 2015 = 100). | State Statistical Office, Price Statistics |
| **Minimum wage** | Log of the value of the minimum wage as stipulated by law. If changes were introduced mid-year, weighted average is taken. | Minimum Wage Law and its amendments |
| **Labor-intensive industries** | A dummy taking a value of 1 for the industry branches 17-22 and 36) and 0 otherwise. | Ranking in Kucera and Sarna (2006) |

## Annex 2 – Robustness analysis with altered variable of central interest

### Table A2.1 – Baseline results

| | FE | IV | | System GMM | |
|---|---|---|---|---|---|
| | | All endogenous | KO and TFP endogenous | All endogenous | KO and TFP endogenous |
| | (1) | (2) | (3) | (4) | (5) |
| **Log labor share lagged** | | | | 0.593*** | 0.637*** |
| | | | | (0.166) | (0.070) |
| **Log capital-output ratio** | 0.218*** | 0.197*** | 0.229*** | 0.0521 | 0.0648 |
| | (0.024) | (0.071) | (0.088) | (0.063) | (0.051) |
| **Log TFP** | -0.0941*** | -0.0606 | -0.121 | -0.108* | -0.188*** |
| | (0.019) | (0.040) | (0.195) | (0.065) | (0.046) |
| **Log prices of inputs** | -0.455 | -0.438 | -0.446 | 0.126 | -0.13 |
| | (0.292) | (0.690) | (0.537) | (0.782) | (0.376) |
| **Log minimum wage** | 0.254 | 0.115 | -0.0652 | 0.840** | 0.409 |
| | (0.156) | (0.239) | (0.385) | (0.419) | (0.295) |
| **Constant** | 1.626 | | | -0.525 | 1.808 |
| | (1.343) | | | (3.805) | (2.030) |
| | | | | | |
| **Observations** | 160 | 120 | 120 | 140 | 140 |
| **Number of cross-sections** | 20 | 20 | 20 | 20 | 20 |
| | | | | | |
| **R-squared** | 66.0% | 26.7% | 18.3% | - | - |
| | | | | | |
| **Hansen test of overid. restrictions (p-value)** Ho: Instruments are valid instruments | - | 0.6814 | 0.3614 | 0.1800 | 0.5930 |
| **Endogeneity test (p-value)** Ho: The specified endogenous regressors can be treated as exogenous | - | 0.9730 | 0.9118 | - | - |

*Source: Authors' estimations.*
**, ** and *** signify a statistical significance at the 10%, 5% and 1% level, respectively. Standard errors given in parentheses. Time dummies included to control for economy-wide shocks over time. Optimal lag length based on the Akaike information criterion, where applicable.*

**Table A2.2 – Disaggregated results, by manufacturing branch**

| | Disaggregation by: | | | |
|---|---|---|---|---|
| | **Capital-output ratio** | **TFP** | **Input prices** | **Minimum wage** |
| | (1) | (2) | (3) | (4) |
| Log capital-output ratio | | 0.234*** | 0.189*** | 0.214*** |
| | | (0.055) | (0.052) | (0.043) |
| Log TFP | -0.0675*** | | -0.0965*** | -0.0876*** |
| | (0.016) | | (0.023) | (0.020) |
| Log prices of inputs | -0.392 | (0.359) | | (0.471) |
| | (0.228) | (0.366) | | (0.303) |
| Log minimum wage | _0.506***_ | 0.247 | 0.129 | |
| | _(0.143)_ | (0.191) | (0.265) | |
| | | | | |
| 15 Food and beverages | 0.191 | -0.0764*** | -1.815*** | 0.595*** |
| | (0.170) | (0.007) | (0.487) | (0.053) |
| 16 Tobacco | _0.729***_ | 0.113*** | (0.320) | -0.701*** |
| | _(0.175)_ | (0.020) | (0.625) | (0.212) |
| 17 Textiles | 0.598*** | 0.110*** | 5.892*** | 6.557*** |
| | (0.161) | (0.005) | (0.342) | (0.288) |
| 18 Wearing apparel | 0.745*** | -0.0631*** | -2.269*** | 1.897*** |
| | (0.159) | (0.014) | (0.307) | (0.484) |
| 19 Leather | _0.662***_ | 0.054 | 1.704*** | _2.289***_ |
| | _(0.174)_ | (0.058) | (0.444) | _(0.706)_ |
| 20 Wood products | -0.053 | -0.0355*** | 2.617*** | 0.193 |
| | (0.190) | (0.010) | (0.551) | (0.137) |
| 21 Paper | -1.237*** | -0.0380** | 3.671*** | -1.681*** |
| | (0.128) | (0.017) | (0.289) | (0.095) |
| 22 Printing | _-0.215_ | (0.003) | 2.110** | _0.335*_ |
| | _(0.206)_ | (0.011) | (0.850) | _(0.163)_ |
| 23 Coke and petroleum | 0.0101 | 0.050 | (0.431) | (0.050) |
| | (0.158) | (0.122) | (1.598) | (0.203) |
| 24 Chemicals | -0.152 | 0.0588*** | 5.788*** | -0.829*** |
| | (0.180) | (0.010) | (1.301) | (0.048) |
| 25 Rubber and plastics | 0.198 | _-0.0219*_ | 2.142*** | _0.327_ |
| | (0.173) | _(0.012)_ | (0.741) | _(0.333)_ |
| 26 Non-metallic products | -0.990*** | (0.001) | 3.760*** | -2.161*** |
| | (0.106) | (0.008) | (0.277) | (0.092) |
| 27 Basic metals | -0.288* | 0.103*** | 0.700 | -1.690*** |
| | (0.157) | (0.025) | (0.505) | (0.224) |
| 28 Fabricated metal products | 0.144 | (0.015) | 1.863*** | -0.923*** |
| | (0.199) | (0.015) | (0.602) | (0.210) |
| 29 Machinery and equipment | _0.421**_ | -0.155*** | 3.644*** | 1.036*** |
| | _(0.148)_ | (0.018) | (0.728) | (0.122) |
| 30 Office and computing machinery | 0.178 | -0.109** | 0.847 | -0.734*** |
| | (0.186) | (0.041) | (0.553) | (0.235) |
| | -0.329** | (0.010) | 1.272** | -2.376*** |
| 31 Electrical machinery and apparatus | (0.153) | (0.006) | (0.571) | (0.189) |
| 32 Media and communication | 0.0501 | 0.050 | 2.990 | -0.238 |
| equipment | (0.177) | (0.065) | (7.960) | (0.391) |
| 35 Other transport | *0.562**** | -0.102*** | 4.639*** | *-0.003* |
| equipment | *(0.147)* | (0.007) | (1.034) | *(0.092)* |
| 36 Furniture | -0.0892 | 0.0428*** | 2.333*** | 0.307 |
| | (0.144) | (0.007) | (0.764) | (0.214) |
| F-test (p-value) Ho: All interactions are jointly significant | 0.0000 | 0.0001 | 0.0000 | 0.0000 |
| Constant | 0.923 | 1.050 | (1.963) | 1.863 |
| | (1.012) | (1.700) | (1.421) | (1.419) |
| Observations | 160 | 160 | 160 | 160 |
| R-squared | 87.9% | 70.3% | 72.8% | 74.1% |
| Number of cross sections | 20 | 20 | 20 | 20 |

*Source: Authors' estimations.*

**, ** and *** signify a statistical significance at the 10%, 5% and 1% level, respectively. Standard errors given in parentheses. Time dummies included to control for economy-wide shocks over time. Optimal lag length based on the Akaike information criterion, where applicable.*